\documentclass[conference]{IEEEtran}
\IEEEoverridecommandlockouts
\usepackage{comment}
\usepackage{cite}
\usepackage{amsmath,amssymb,amsfonts}
\usepackage{algorithmic}
\usepackage{graphicx}
\usepackage{textcomp}
\usepackage{xcolor}
\usepackage{hyperref}
\usepackage{booktabs}
\usepackage{balance}
\def\BibTeX{{\rm B\kern-.05em{\sc i\kern-.025em b}\kern-.08em
    T\kern-.1667em\lower.7ex\hbox{E}\kern-.125emX}}

\begin{document}

\title{Text Prompt is Not Enough: Sound Event Enhanced Prompt Adapter for Target Style Audio Generation}

\author{
    \IEEEauthorblockN{Chenxu Xiong$^{1}$, Ruibo Fu$^{2}$\IEEEauthorrefmark{1}\thanks{*Corresponding author}, Shuchen Shi${^2}$, Zhengqi Wen$^{3}$, Jianhua Tao$^{3}$, Tao Wang$^{2}$, Chenxing Li$^{4}$, \\ Chunyu Qiang$^{2}$, Yuankun Xie$^{2}$, Xin Qi$^{2}$, Guanjun Li$^{2}$, Zizheng Yang$^{1}$}
    \IEEEauthorblockA{$^1$ SDU-ANU Joint Science College, Shandong University, Weihai, China }
    \IEEEauthorblockA{$^2$ Institute of Automation, Chinese Academy of Sciences, Beijing, China}
    \IEEEauthorblockA{$^3$ Department of Automation, Tsinghua University, Beijing, China}
    \IEEEauthorblockA{$^4$ AI Lab, Tencent, Beijing, China}
    \IEEEauthorblockA{202100700062@mail.sdu.edu.cn, ruibo.fu@nlpr.ia.ac.cn}
    
}

\maketitle

\begin{abstract}
Current mainstream audio generation methods primarily rely on simple text prompts, often failing to capture the nuanced details necessary for multi-style audio generation. To address this limitation, the Sound Event Enhanced Prompt Adapter is proposed. Unlike traditional static global style transfer, this method extracts style embedding through cross-attention between text and reference audio for adaptive style control. Adaptive layer normalization is then utilized to enhance the model's capacity to express multiple styles. Additionally, the Sound Event Reference Style Transfer Dataset (SERST) is introduced for the proposed target style audio generation task, enabling dual-prompt audio generation using both text and audio references. Experimental results demonstrate the robustness of the model, achieving state-of-the-art Fréchet Distance of 26.94 and KL Divergence of 1.82, surpassing Tango, AudioLDM, and AudioGen. Furthermore, the generated audio shows high similarity to its corresponding audio reference. The demo, code, and dataset are publicly available.\footnote{\href{https://michael1223132.github.io/PromptAdapter/}{https://michael1223132.github.io/PromptAdapter/}}
\end{abstract}

\section{Introduction}
Target Style Audio Generation generates audio with specific styles or features, allowing for more natural and fine-grained audio production. This approach has numerous applications, particularly in the media industries, where it can generate background sound effects that match specific scenes. The current mainstream method for audio generation is Text-to-Audio (TTA) \cite{yang2023diffsound,kreuk2022audiogen,huang2023makeanaudio,liu2023audioldm,ghosal2023tango}. These TTA models, often encoded by CLAP \cite{elizalde2022claplearningaudioconcepts} or T5 \cite{Raffel2019ExploringTL}, utilize rich semantic information in textual descriptions to produce high-quality audio outputs.

Although mainstream methods using single-text prompts have achieved promising results, several limitations remain. Text input and audio output belong to different modalities, making alignment between the two challenging. From a mathematical perspective, achieving full control over the generated audio requires the mapping between input and output to be at least surjective, if not bijective. For instance, generating the sound of a dog barking from a single text prompt fails to capture specific characteristics such as timbre or how the environment interacts with the barking. This limitation restricts the ability to model audio in finer detail. To address this issue, incorporating additional prior knowledge is essential for providing richer contextual information and enhancing the precision of the generated output.

Two primary approaches exist for introducing prior knowledge into audio generation. The first involves control conditions manipulating the generated audio's pitch, energy, and temporal relationships \cite{Guo2023AudioGW, Xie2024PicoAudioEP, Liao2024BATONAT}. However, no current methods specifically address style control in audio generation. The second approach utilizes multi-modal prompts that incorporate semantic and temporal information from other modalities, such as images \cite{Sheffer2022IHY} and videos \cite{SpecVQGAN_Iashin_2021, luo2023difffoley, xu2024vta-ldm}. Despite their potential, cross-modal prompts often suffer from interference caused by redundant and unrelated information, as they do not provide intuitive acoustic references for the model. As a result, a text-acoustic fusion prompt emerges as an effective solution, not only providing intuitive information to the model but also filling the gap in style control.

\begin{figure}[t]
\centering
\includegraphics[width=0.5\textwidth]{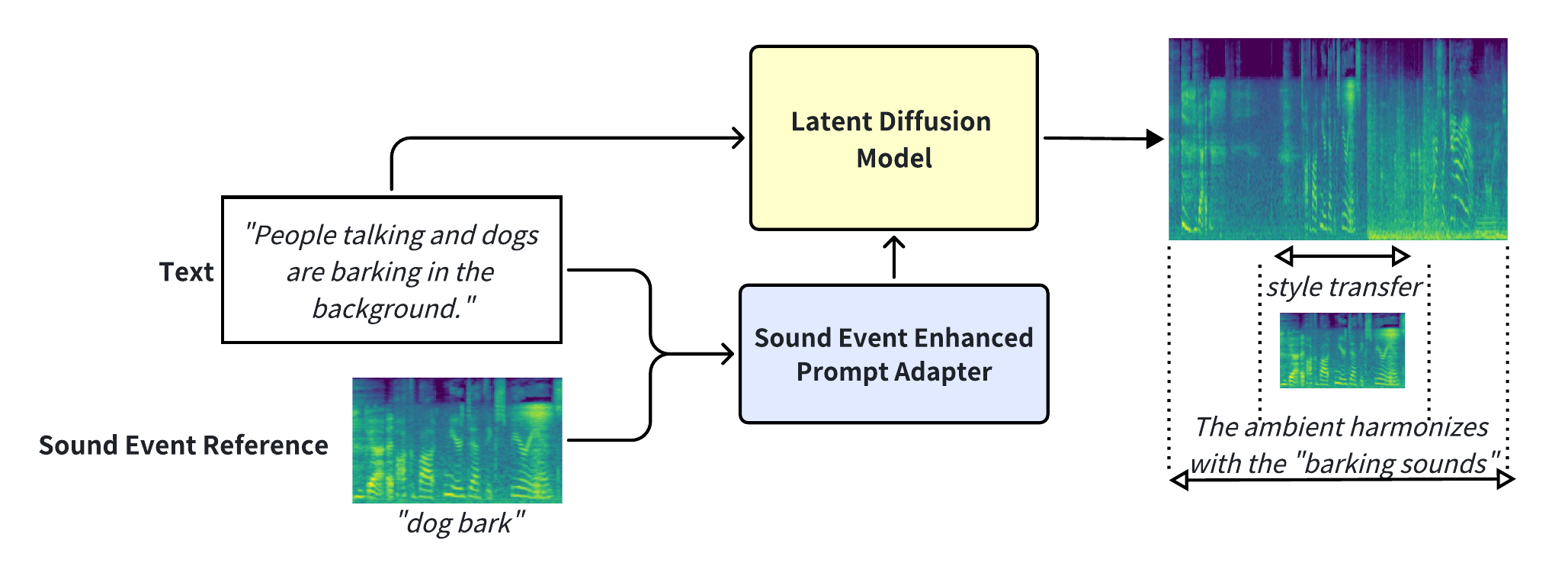}
\caption{Example of Sound Event Enhanced Prompt Adapter, generating audio while preserving the style from sound event.}
\label{usage}
\end{figure}

\begin{figure*}[htbp]
\centering
\includegraphics[width=\textwidth]{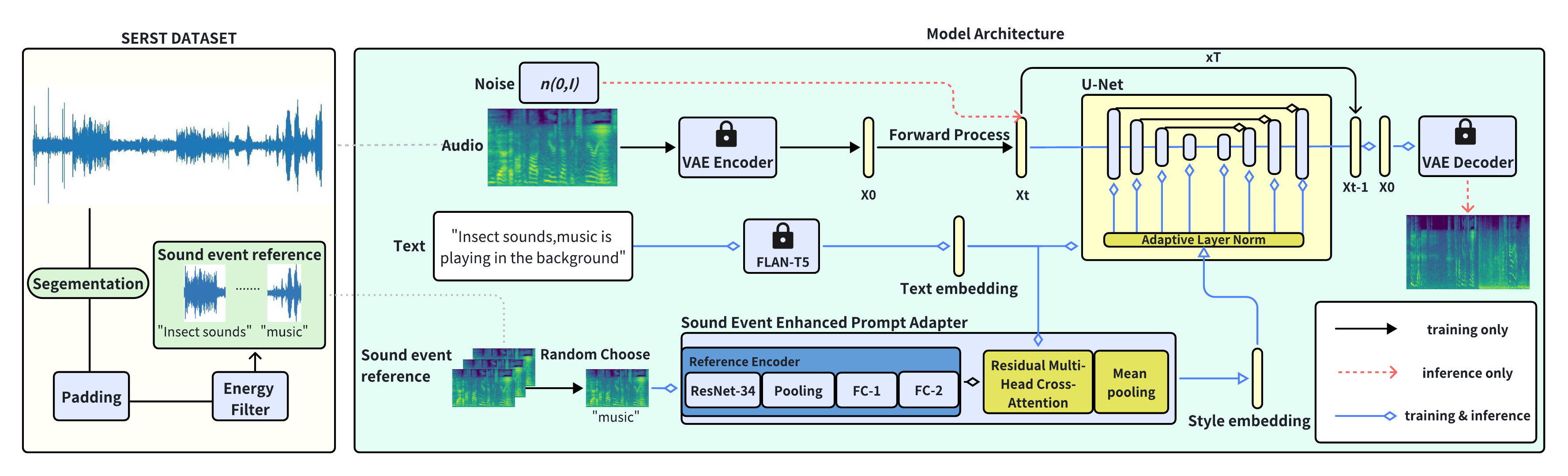}
\caption{SEARST Dataset and Model Architecture: In the training stage, the latent diffusion model (LDM) is conditioned on an audio embedding learned in a continuous space through a variational auto-encoder (VAE). Text is fused with a randomly selected sound event reference through the Sound Event Enhanced Prompt Adapter to generate a style embedding. This style embedding is then utilized for adaptive layer normalization in the U-Net. In the inference stage, the LDM is conditioned on random noise instead of the audio embedding derived from the VAE. }
\label{architecture}
\end{figure*}

In this paper, we first propose the Sound Event Enhanced Prompt Adapter. Traditional style transfer approaches typically extract a global style directly from the reference. However, text offers valuable semantic information that can guide and refine the application of this global style. To leverage this, cross-attention \cite{Vaswani2017AttentionIA} is employed between sound events and text to identify which text events are most closely correlated with the corresponding audio reference, as illustrated in Fig. \ref{usage}. Additionally, the style embedding generated by the adapter is passed into the U-Net \cite{Ronneberger2015UNetCN} via adaptive layer normalization \cite{Peebles2022ScalableDM}, which allows the normalization layer to adapt to the data distribution from style embedding. We then construct a Sound Event Reference Style Transfer Dataset (SERST) that integrates dual-modality prompts from event-level audio reference and text, derived from Audioset-Strong \cite{hershey2021temporal}. Experimental results demonstrate the robustness of the proposed method across various sound event references, and significant improvements in the accuracy of acoustic modeling. Specifically, the method achieves gains of 2.3\% in Fréchet Distance and 7.6\% in KL divergence. Additionally, the generated audio exhibits a strong alignment with its audio reference, as indicated by a score of 0.4 in CLAP-audio similarity. The key
contributions of this paper are summarized as follows:

\begin{enumerate}
\item A new audio generation task is introduced, guided by both text and sound event references, enabling the transfer of style from the reference and improving the accuracy and naturalness of audio generation.
\item A new dataset, SERST, is created by integrating existing datasets, consisting of audio and sound event segments. Evaluation metrics were applied to assess performance, providing a benchmark for future research.
\item A Sound Event Enhanced Prompt Adapter is proposed that adaptively transfers the style from reference audio through cross-attention between the text and reference audio segments, integrated with an adaptive layer normalization within the U-Net. This approach enables finer-grained control over the audio generation process that enhances the accuracy of acoustic modeling and achieves target style transfer. 
\end{enumerate}

\section{Method}
To achieve target style audio generation, a Sound Event Reference Style Transfer Dataset is built. We utilize the Sound Event Enhanced Prompt Adapter to extract style embedding, which is then sent into the Conditional Latent Diffusion Audio Generation Model. It comprises a Variational Auto-Encoder (VAE) \cite{Kingma2013AutoEncodingVB}, a conditional Latent Diffusion Model (LDM) \cite{Rombach2021HighResolutionIS}, and a text condition encoder (FLAN-T5) \cite{Chung2022ScalingIL}. The latent representation constructed by LDM is then used to generate the mel-spectrogram via the VAE decoder. A vocoder is employed to generate the audio in the inference phase. The overall architecture is illustrated in Fig. \ref{architecture}.

\subsection{Sound Event Reference Style Transfer Dataset (SERST)}
Effective style transfer requires high-quality reference audio. To address this need, the Sound Event Reference Style Transfer Dataset (SERST) is constructed, providing event-level granularity audio to capture the full distribution of acoustic events and enabling the accurate reconstruction of their characteristics. This dataset is created by segmenting the original audio from the Audioset-Strong dataset \cite{hershey2021temporal} based on annotated acoustic event timestamps. Statistical analysis revealed that a 2-second audio length offers an optimal balance between segment quantity and accuracy. Audio is segmented by event, and in cases where the segments are shorter than 2 seconds, they are concatenated from other clips with the same sound event tag, facilitating both padding and data augmentation. Then a Short-Time Energy detection is used to filter out Poor quality references. As a single original audio could yield multiple trimmed segments: during training, one of these segments is randomly selected, while during inference, all trimmed segments are utilized to examine the variability in the generated audio. The dataset consists of 88,464 training samples, 1,384 validation samples, and 1,180 test samples.

\subsection{Sound Event Enhanced Prompt Adapter}
To fully utilize the acoustic information, the global sound event style feature is extracted from a reference encoder. A style embedding is then generated through cross-attention between the text and reference audio, enabling adaptive style transfer and allowing the model to focus on the relevant aspects of the reference audio’s style.

The sound event reference is first compressed into a reference embedding $e_{\text{r}}$, representing the global style of the audio. Given the lack of suitable pre-trained encoders for this task, a custom reference audio encoder was developed based on the H/ASP model \cite{Heo2020ClovaBS}, originally designed for Text-to-Speech (TTS). This encoder consists of a ResNet-34 backbone, a pooling layer, and two fully connected layers, designed to capture the fine-grained acoustic features required for this task. The global style is then integrated with local information from the text condition $e_{\text{t}}$. Residual cross-attention between the text embedding and the audio embedding is applied to generate the style embedding $e_{s}$:
\begin{align}
    Q &= e_{t} W_q, \quad K = e_r W_{k}, \quad V = e_{r} W_v, \\
    e_{s} &= \text{Softmax} \left( \frac{Q K^T}{\sqrt{d/h}} \right) \cdot V + e_{t}
\end{align}
Where $d$ represents the embedding dimension of $e_{\text{t}}$, and $h$ refers to the number of multi-heads. We then perform mean pooling along the sequence length dimension to align dimensions and feed them into U-Net.

\subsection{Conditional Latent Diffusion Audio Generation Model}
The latent diffusion model (LDM) constructs the audio prior $z_{0}$ with the guidance of text and audio. LDM can achieve this through a forward and reverse diffusion process. The forward diffusion is a Markov chain of Gaussian distributions with scheduled noise parameters $0 < \beta_1 < \beta_2 < \cdots < \beta_N < 1$. Through a reparametrization trick that allows direct sampling of any $z_{n}$ from $z_{0}$ via a non-Markovian process: 
\begin{equation}
z_n = \sqrt{\bar{\alpha}_n} z_0 + (1 - \bar{\alpha}_n) \epsilon,
\end{equation}
where $\epsilon$ is a standard Gaussian noise and ${\bar{\alpha}_n} = \prod_{s=1}^{t} (1 - \beta_s)$ . The LDM model aims to conduct the denoising process on mel-embedding (training) or standard Gaussian noise $\epsilon$ (inference) and predict the mel-embedding $\hat{x_{0}}$. For every step t, the training objective is to minimize the following: 
\begin{equation}
    \mathcal{L}_{\text{LDM}} = \mathbb{E}_{x, \epsilon \sim \mathcal{N}(0, I)} \left\| \epsilon_\theta(x_t, t, e_{\text{t}}, e_{\text{r}}) - \epsilon \right\|_2^2.
\end{equation}
In this context, $\epsilon_{\theta}$ represents the noise estimation conditioned on $t$, $e_{t}$ and $e_{r}$. The architecture of the LDM primarily utilizes a U-Net structure \cite{Ronneberger2015UNetCN}, which consists of a series of ResNet\cite{He2015DeepRL} and transformer blocks.

 The shift parameters $\gamma$ and $\beta$, derived from the concat of style embedding and time step embedding, are applied as adaptive layer normalization-zero parameters \cite{Peebles2022ScalableDM} throughout the Resnet blocks in U-Net. This is because the adaptive layer norm allows the normalization layer to adapt to data distributions in different modalities or domains, thus performing well in multimodal learning or domain adaptation tasks. This dynamic adjustment is achieved by learning how to modify the normalized mean and variance based on data, thereby generating more robust feature representations.

To guide the reverse diffusion process to reconstruct the audio prior $z_{0}$, we employ a classifier-free guidance \cite{Ho2022ClassifierFreeDG} of condition input $\tau$. During training, the guidance was randomly dropped for 10\% of the training samples. When inference, a guidance scale $w$ controls the contribution of guidance to the noise estimation $\hat{\epsilon}_\theta$, in comparison to unguided estimation where empty text is passed:
\begin{equation}
\hat{\epsilon}_\theta^{(n)}(z_n, \tau) = w \epsilon_\theta^{(n)}(z_n, \tau) + (1 - w)\epsilon_\theta^{(n)}(z_n).
\end{equation}

\section{Experiments}

\subsection{Training Setting}\label{AA}
All data were resampled to a 16kHz sampling rate, with each sample padded to a duration of 10.24 seconds. The VAE and text condition encoder were kept frozen and accepted audio at 16kHz while we fine-tuned the latent diffusion model using pre-trained weights from Tango \cite{ghosal2023tango}. The reference audio encoder, in contrast, was trained from scratch. The text encoder is based on FLAN-T5-LARGE \cite{Chung2022ScalingIL}, which contains a total of 780 million parameters. HiFi-GAN \cite{kong2020hifigan} was used as the vocoder to convert mel-spectrograms into audio. The trainable components include the U-Net, which loaded the pre-trained weights from Tango, and the reference audio encoder, collectively comprising 1.097 billion trainable parameters. We employed AdaFactor as the optimizer and AdafactorSchedule as the scheduler to accelerate the training process. Our model was trained for 20 epochs on four RTX 3090 GPUs with a batch size of two. The checkpoint with the lowest validation loss was then selected for final evaluation.

\subsection{Evaluation Metrics}\label{AA}

We compared our model to Tango\cite{ghosal2023tango}, AudioGen \cite{kreuk2022audiogen} and AudioLDM\cite{liu2023audioldm} and used four objective metrics: Fréchet Distance (\textbf{FD}), Fréchet Audio Distance (\textbf{FAD}), KL divergence (\textbf{KL}), Mel-Spectrogram cosine Similarity (\textbf{Mel-Sim}) and Clap-Audio\cite{elizalde2022claplearningaudioconcepts} cosine similarity (\textbf{CLAP-Audio}). The first two measure the distance between the generated audio distribution and the real audio distribution while the third one computes the divergence between the distributions of the original and generated audio samples. To calculate the Mel-Spectrogram cosine similarity between the sound event reference and generated audio, we segmented the generated audio into multiple parts. We calculated the similarity for each segment against the reference audio. The highest similarity value among these segments was then taken as the overall similarity between the generated and reference audio. 
The CLAP-Audio cosine similarity is employed to measure the similarity between different generated audios and between the generated audios and their respective references.

As for subjective evaluation, we paid twenty experienced human evaluators to assess fifty randomly selected audio samples on a scale from 1 to 100 in the following aspects: overall audio quality (\textbf{OVL}) and relevance to the input text (\textbf{REL}) that reflects the quality of generated audio and its relevance to the input sound event prompt (\textbf{REA}) that demonstrates the ability in target style transfer. 
\begin{table}[htbp]
\caption{Sensitivity Analysis. The results show the Clap similarity of our generated audios under identical sound event reference (ID ref) or different sound (Diff ref) event reference.}
\label{tab:sensitivity}
\begin{center}
\begin{tabular}{|c|c|c|c|}
\hline
\textbf{Ours} & \textbf{\textit{ID ref}}& \textbf{\textit{Diff ref}}& \textbf{\textit{Diff}} \\
\hline
CLAP-Audio& 0.72& 0.54& \textbf{0.18}  \\
\hline
\end{tabular}
\label{tab1}
\end{center}
\end{table}

\begin{table}[t]
\centering
\caption{Model effectiveness. The results of the model effectiveness show the accuracy of generated audio from Our model compared to different baseline models.}
\label{tab:compare}
\begin{tabular}{lccc|cc}
\toprule
\textbf{Models} & \multicolumn{3}{c|}{\textbf{Objective Metrics}} & \multicolumn{2}{c}{\textbf{Subjective Metrics}} \\
\cmidrule(lr){2-4} \cmidrule(lr){5-6}
 & \textbf{FD \textcolor{green} {$\downarrow$}} & \textbf{FAD \textcolor{green}{$\downarrow$}} & \textbf{KL \textcolor{green}{$\downarrow$}} & 
 \textbf{OVL \textcolor{red}{$\uparrow$}} & \textbf{REL \textcolor{red}{$\uparrow$}} \\
\midrule
Ground truth & -- & -- & -- &87.50 & 83.65 \\
\midrule
AudioGen \cite{kreuk2022audiogen} & 28.52 & 2.47 & 2.12  &73.25 & 71.90 \\
AudioLDM \cite{liu2023audioldm}& 28.07 & 2.44 & 2.01  &72.60 & 69.85 \\
Tango \cite{ghosal2023tango}& 27.60 & \textbf{2.21} & 1.97 & 74.40 & 75.40 \\
\midrule
\textbf{Ours} & \textbf{26.94} & 2.38 & \textbf{1.82} & \textbf{79.10} & \textbf{77.65} \\
\bottomrule
\end{tabular}
\vspace{0.1cm}
\end{table}

\begin{table}[t]
\centering
\caption{Ablation study. The results of the ablation experiment showing the performance of different fusion methods in terms of objective metrics. The input channel implies where the style embedding will be sent into U-net after fusion. Fusion type means how we fuse text with reference.}
\label{tab:ablation}
\resizebox{0.48\textwidth}{!}{
\begin{tabular}{lcc|ccc}
\toprule
\textbf{Model} & \multicolumn{2}{c|}{\textbf{Fusion Method}} & \multicolumn{3}{c}{\textbf{Objective Metrics}} \\
\cmidrule(lr){4-6} 
 & \textbf{Input Channel} & \textbf{Fusion Type} & \textbf{FD \textcolor{green}{$\downarrow$}} & \textbf{FAD \textcolor{green}{$\downarrow$}} & \textbf{KL \textcolor{green}{$\downarrow$}} \\
\midrule
\textbf{Ours} & \textbf{Timestep} & \textbf{Cross Attention} & \textbf{26.94} & \textbf{2.38} & \textbf{1.88} \\
Variant1 & Timestep & Concat & 28.54 & 3.14 & 1.93 \\
\midrule
Variant2 & Text & Cross Attention & 39.15 & 6.09 & 2.27 \\
Variant3 & Text & Concat & 38.50 & 4.35 & 2.30 \\
\bottomrule
\end{tabular}
}
\vspace{0.1cm}

\end{table}

\begin{table}[t]
\centering
\caption{Audio Relevance Evaluation. The results emphasize the alignment between the generated audio and its reference.}
\label{tab:audio_relevance}
\resizebox{0.4\textwidth}{!}{
\begin{tabular}{lccc}
\toprule
\textbf{Models} & \textbf{Mel-Sim \textcolor{red}{$\uparrow$}} & \textbf{CLAP-Audio\textcolor{red}{$\uparrow$}} & \textbf{REA \textcolor{red}{$\uparrow$}} \\
\midrule
AudioGen \cite{kreuk2022audiogen} & 0.71 & 0.33 & 63.35 \\
AudioLDM \cite{liu2023audioldm} & 0.70 & 0.32 & 64.00 \\
Tango \cite{ghosal2023tango} & 0.71 & 0.34 & 64.25 \\
\midrule
Variant1 & 0.73 & 0.36 & 69.00 \\
\textbf{Ours} & \textbf{0.76} & \textbf{0.40} & \textbf{76.00} \\
\bottomrule
\end{tabular}
}
\vspace{0.1cm}
\end{table}

\section{Results and Analysis}
In this section, we first conduct a sensitivity analysis on the Sound Enhanced Prompt Adapter to evaluate its effectiveness. Then grade its performance (referred to as \textbf{Ours}) in comparison to baseline models. Additionally, we conduct an ablation study on our model and various modified versions, focusing on identifying the most effective method for fusing the modalities. Finally, we assess the relevance of the generated outputs and the provided reference audio.

\subsubsection{\textbf{Sensitivity Analysis for Sound Enhanced Prompt Adapter}}
Table \ref{tab:sensitivity} presents the CLAP-Audio similarity results of the generated audio provided with various sound event references, while keeping the text input constant. When the same sound event reference is provided to the model multiple times, the generated audio exhibits a CLAP similarity score of 0.72. In contrast, when different sound event references are used, the generated outputs yield a CLAP similarity score of 0.54. This difference of 0.18 demonstrates the effectiveness of the Sound Enhanced Prompt Adapter in utilizing prior acoustic information.

\subsubsection{\textbf{Comparison of Generated Audio Accuracy with Baseline Models}}
Table \ref{tab:compare} presents the evaluation results of our model compared to TTA models using both objective and subjective metrics. Our model achieves great results in both objective and subjective evaluation. In terms of objective metrics, our model achieves an FD score of 26.94, and a KL divergence of 1.88, which are all the lowest in all models. The FAD score of 2.38, although not the best, is still very competitive and close to Tango’s leading result of 2.21. For subjective metrics, our model achieves an overall quality (OVL) score of 79.10 and a relevance (REL) score of 77.65, which are both the best in these models, showing that the audio generated by our model is very well aligned with the provided textual descriptions.

\subsubsection{\textbf{Ablation Study of Text and Sound Event Prompt Fusion Methods}}

Table \ref{tab:ablation} presents the results of our ablation study. We experimented with four different approaches: concatenating the reference embedding with the text embedding or applying cross-attention to obtain the style embedding, then sending the merged style embedding into U-net either with the text input or integrating it into the layer normalization of ResNet blocks within the U-Net, alongside the timestep embedding. The results indicate that using cross-attention to generate the style embedding, followed by its incorporation into the layer normalization, yields the best performance.

\subsubsection{\textbf{Style transfer performance evaluation by measuring audio similarity}}

Tabel \ref{tab:audio_relevance} presents the evaluation results for the similarity of generated audio and sound event reference. Our model achieves the highest scores in all metrics, with a Mel-Sim of 0.76, CLAP-Audio similarity of 0.40, and an REA of 76.00, demonstrating strong relevance with the reference compared to the other models. While Variant1 produces respectable results, it falls short of our model’s performance. In contrast, AudioGen, AudioLDM and Tango show lower scores. These results underscore the effectiveness of our approach in leveraging sound event reference to transfer the generated audio.

\section{Conclusion}

This work first introduces the SERST dataset, which integrates dual-modality prompts from event-level audio reference and text, providing a valuable resource for target audio generation. Then a Sound Event Enhanced Prompt Adapter is proposed to achieve fine-grained style control in audio generation. The method leverages cross-attention and adaptive layer normalization, significantly improving the quality and controllability of generated audio, particularly in style. Compared to Tango, the proposed approach improves FD and KL Divergence scores by 2.3\% and 7.6\%. The generated audio strongly aligns with the reference audio, highlighting effective style control. Future work will explore additional methods to enhance the performance of the prompt adapter.

\balance
\bibliographystyle{IEEEtran}
\bibliography{main}

\vspace{12pt}
\color{red}

\end{document}